\documentclass{article}
\usepackage{spconf,amsmath,amssymb,graphicx,booktabs}
\usepackage[table]{xcolor}
\usepackage[hidelinks]{hyperref}

\title{WHERE SPEECH ENHANCEMENT HURTS RECOGNITION:\\
AN INFERENCE-TIME POLAR PROJECTION DIAGNOSIS}
\name{Mingyue Huo$^{1}$ \qquad Yuheng Zhang$^{1}$\qquad Hao Zhang$^{2}$ }
\address{$^{1}$University of Illinois Urbana-Champaign \quad
$^{2}$Wuhan University\\[-0.1em]
{\footnotesize\texttt{mhuo5@illinois.edu} \quad \texttt{yuhengz2@illinois.edu} \quad \texttt{h.zhangnwpu@gmail.com}}}

\makeatletter
\def\@maketitle{\newpage
 \null
 \vskip -1em \begin{center}
 {\large \bf \@title \par} \vskip 1.0em {\large \lineskip .35em
\begin{tabular}[t]{c}\@name \\ \@address
\end{tabular}\par}
 \end{center}
 \par
 \vskip 1.0em}
\makeatother

\begin{document}
\ninept
\maketitle

\begin{abstract}
Speech enhancement (SE) can substantially improve perceptual quality, yet
enhanced speech does not necessarily improve automatic speech recognition
(ASR). Common explanations such as enhancement artifacts and over-suppression
remain qualitative and do not localize which enhancement component affects
recognition. We study \emph{inference-time polar projection}, which transforms
an STFT mask $M=Ae^{j\phi}$ into $M_{\alpha,\gamma}=A^\alpha
e^{j\gamma\phi}$ and independently varies magnitude strength and estimated
phase correction with frozen SE and ASR models. The resulting response curves
separate effects that are coupled in ordinary enhanced-vs-noisy comparisons.
On VoiceBank+DEMAND with FRCRN, wav2vec~2.0 improves from 17.13 to 10.20 WER
near full magnitude correction ($\alpha=.85$), while Whisper reaches its best
WER under much weaker correction, improving from 6.44 to 5.57 at
$\alpha=.25$ and returning to 6.46 at full magnitude. A second enhancer, a
robustness-oriented ASR model, and a reverberant WSJ0 mixture set generalize the pattern and show that the
preferred magnitude region depends on the recognizer and acoustic condition.
Estimated phase correction is not a stable ASR lever in these evaluations: its
effect is smaller, setting-dependent, and provides no consistent gain once
magnitude strength is fixed.
\end{abstract}


\section{Introduction}
\label{sec:intro}

Speech enhancement (SE) has made substantial progress in improving signal
quality and intelligibility for human listeners. Modern enhancers no longer
only suppress noise in the magnitude spectrum; they also model phase, either
explicitly through complex ratio masks~\cite{williamson2016crm,zhao2022frcrn}
and phase reconstruction~\cite{wang2018end}, or implicitly through time-domain
waveform reconstruction~\cite{rethage2018wavenet,luo2019conv}. Other systems
further optimize perceptual objectives such as PESQ~\cite{fu2021metricgan+}.
These advances are well motivated for human listening, where both magnitude
and phase can affect perceived quality~\cite{paliwal2003usefulness,gerkmann2015phase}.

However, better enhanced speech does not necessarily help a machine listener.
Recent studies report that SE front ends can yield only marginal ASR gains, and
in some cases even higher word error rate (WER) than the unprocessed noisy
input~\cite{iwamoto2022artifacts,ochiai2024rethinking}. This mismatch also
persists for modern recognizers such as wav2vec~2.0 and Whisper, which differ
substantially in input representation and robustness. The degradation is often
attributed to enhancement artifacts or to \emph{over-suppression}, where the
enhancer removes low-energy speech cues that ASR still relies on
~\cite{braun2022effect}. Yet these
explanations are usually qualitative. They do not specify whether the harmful
component is magnitude modification, phase correction, or their interaction,
nor whether the same failure mode holds across ASR architectures.

Existing remedies improve ASR performance but do not fully answer this
diagnostic question. One line of work retrains the enhancer with ASR-friendly
objectives or jointly optimizes the SE and ASR models~\cite{wang2016joint}.
These approaches can be effective, but require model retraining and become
expensive when the downstream recognizer is large or fixed. Another line
performs inference-time post-processing, such as interpolating enhanced speech
with the noisy observation to reintroduce information removed by enhancement
\cite{iwamoto2022artifacts,sato2022learning}. Such methods are lightweight,
but adjust multiple enhancement effects together and therefore cannot reveal
which component is associated with the ASR change.

We introduce \emph{inference-time polar projection} for diagnosing
SE-induced ASR degradation. Given a complex or effective mask $M=Ae^{j\phi}$,
we form $M_{\alpha,\gamma}=A^\alpha e^{j\gamma\phi}$, where $\alpha$ controls
magnitude strength and $\gamma$ controls phase correction. Sweeping these two
parameters with frozen SE and ASR models turns qualitative explanations into
measurable response curves. Our contributions are threefold:
\begin{itemize}
\item We introduce an inference-time diagnostic with independently settable
magnitude- and phase-strength controls. Unlike a one-knob strength adjustment,
the probe distinguishes the two mask-level operations.
\item We reveal a pronounced recognizer dependence in the magnitude response.
Estimated-phase interventions have smaller, setting-dependent effects and
provide no consistent ASR benefit over the evaluated slices.
\item We test the pattern with two architecturally distinct enhancers, three
recognizer checkpoints, and two acoustic test sets. These comparisons broaden
the evidence without attributing the behavior to a single model property.
\end{itemize}

\begin{table*}[t]
\caption{Interpreting polar projection. Each row is an inference-time mask
operation on the same frozen enhancer output. The Setting column gives the
corresponding $(\alpha,\gamma)$ values when applicable; observation adding is
included as a coupled one-knob contrast.}
\label{tab:interpretation}
\centering
\footnotesize
\setlength{\tabcolsep}{3.2pt}
\renewcommand{\arraystretch}{1.12}
\begin{tabular}{@{}p{.14\textwidth}p{.18\textwidth}p{.11\textwidth}p{.50\textwidth}@{}}
\toprule
Case & Mask & Setting & Interpretation\\
\midrule
Identity & $1$ & $(0,0)$ & No modification; noisy input.\\
Complex mask & $Ae^{j\phi}$ & $(1,1)$ & Native complex-mask output~\cite{williamson2016crm,zhao2022frcrn}.\\
Magnitude mask & $A$ & $(1,0)$ & Learned magnitude with mixture phase; related to magnitude anchoring~\cite{wang2021compensation}.\\
Magnitude strength & $A^\alpha$ & $(\alpha,0)$ & One-axis mask weakening, related to mask scalar or warping controls~\cite{narayanan2022maskscalar,wang2021warping}.\\
Phase strength & $A e^{j\gamma\phi}$ & $(1,\gamma)$ & Full magnitude with scaled estimated phase.\\
Polar projection & $A^\alpha e^{j\gamma\phi}$ & $(\alpha,\gamma)$ & Independent magnitude and phase strengths (ours).\\
Observation adding & $\lambda M+(1-\lambda)$ & -- & Coupled interpolation of magnitude, phase, energy, and artifacts~\cite{iwamoto2022artifacts,sato2022learning}.\\
\bottomrule
\end{tabular}
\end{table*}

\section{Related work}
\label{sec:related}

Speech enhancement is a natural ASR front end, but improved perceptual quality
does not always lower recognition error. Studies analyzing the degradation
from the consequence side decompose enhancement errors into interference,
noise, and artifact components to which ASR is most sensitive
\cite{iwamoto2022artifacts,ochiai2024rethinking}. These explanations remain
output-level: they identify what kind of distortion hurts ASR, not which
mask-level operation, magnitude modification or phase correction, is
associated with it.

Training-based methods change SE and/or ASR parameters so enhancement serves
recognition. Examples include end-to-end joint training with an ASR objective,
ASR-supervised adaptive mask scalars, and fused noisy--enhanced streams
\cite{wang2016joint,narayanan2022maskscalar,pandey2021dual,hu2022interactive}.
Magnitude--phase compensation during SE training can also affect ASR
\cite{wang2021compensation}. Inference-time methods keep the
models fixed and edit the output. Observation adding interpolates enhanced and
noisy signals to trade residual noise against artifacts
\cite{iwamoto2022artifacts,sato2022learning}; others soften a mask through
exponents or task-aware warping~\cite{wang2021warping}. Our polar projection builds
on this strength-control literature rather than claiming magnitude scaling as
new. Its distinction is diagnostic: it intervenes on the enhancer's STFT mask,
separating magnitude strength from phase correction and shifting the question
from whether weaker enhancement helps to which mask-level operation tracks the
response.

\section{Method}
\label{sec:method}

\begin{figure}[t]
  \centering
  \includegraphics[width=.60\columnwidth,trim=40 45 40 15,clip]{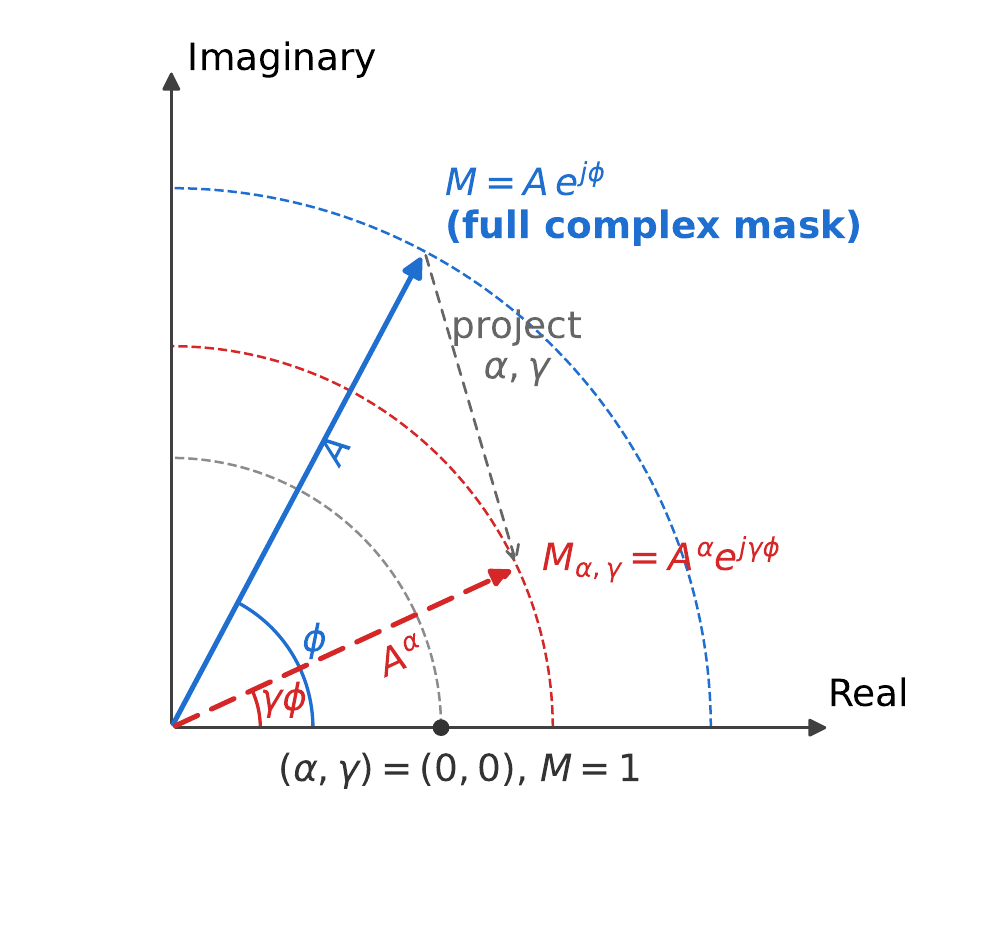}
  \vspace{-10pt}
  \caption{Polar projection in the complex-mask plane. Magnitude strength
  $\alpha$ and phase-correction strength $\gamma$ move the mask toward the
  identity along independently settable axes.}
  \label{fig:projection}
\end{figure}

Let $X(t,f)\in\mathbb{C}$ denote the mixture STFT. A mask-based enhancer
estimates $M=M_r+jM_i$, and reconstructs $\hat S=MX$. Writing
$M=Ae^{j\phi}$, where $A=(M_r^2+M_i^2)^{1/2}$ and
$\phi=\operatorname{atan2}(M_i,M_r)$, separates the magnitude scaling from the
phase correction relative to the mixture. Writing the mixture as
$X=|X|e^{j\theta_X}$ gives
$\hat S=A|X|e^{j(\theta_X+\phi)}$: full complex-mask enhancement applies both a
magnitude modification and a phase correction. To diagnose their effects
separately, we replace the mask at inference by
\begin{equation}
 M_{\alpha,\gamma}=A^\alpha e^{j\gamma\phi},\qquad
 \hat S_{\alpha,\gamma}=M_{\alpha,\gamma}X,
 \quad \alpha,\gamma\in[0,1].
 \label{eq:projection}
\end{equation}
Table~\ref{tab:interpretation} lists the operating points used in this
parameterization. The pair $(\alpha,\gamma)$ specifies how much of the
predicted mask magnitude and phase correction is retained: $(0,0)$ is the
identity mask, $(1,0)$ keeps only the predicted magnitude, and $(1,1)$ recovers
the full complex-mask output for a strict mask enhancer. With $\gamma=0$,
$\log|\hat S_{\alpha,0}|=\log|X|+\alpha\log A$, so $\alpha$ is a linear
coordinate along the enhancer's predicted log-magnitude correction. We limit
both controls to the path between the identity and predicted mask; values above
one, which would over-apply the correction, are outside this study. The
controls are independently settable, but Eq.~\eqref{eq:projection} does not
assume their ASR effects are additive. Figure~\ref{fig:projection} illustrates
the two paths.

\begin{figure*}[t!]
  \centering
  \includegraphics[width=0.97\textwidth]{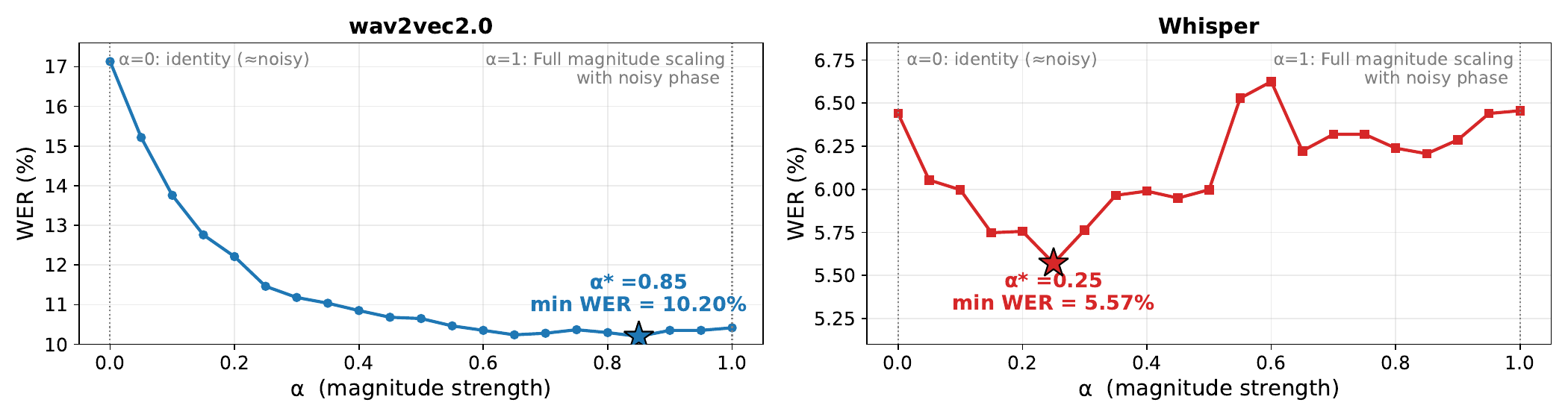}
  \vspace{-10pt}
  \caption{FRCRN magnitude sweep at $\gamma=0$. wav2vec~2.0 improves sharply
  and then reaches a broad plateau; Whisper attains its observed minimum under
  a weaker correction. Stars mark evaluation-grid minima, not selected
  deployment settings.}
  \vspace{-10pt}
  \label{fig:alpha}
\end{figure*}

Observation adding is a common inference-time way to improve ASR after
enhancement by mixing the enhanced signal with the observation. In our notation
it uses $\tilde x_\lambda=\lambda\hat x+(1-\lambda)x$. By iSTFT linearity,
this applies $M_\lambda=\lambda M+(1-\lambda)\mathbf 1$ and jointly changes
magnitude, phase, energy, artifacts, and STFT-consistency effects. Polar
projection targets a different question: it does not introduce a new enhancer,
but uses the mask factorization as a controlled probe before waveform
reconstruction. A recognizer then re-analyzes that waveform, so varying
$\gamma$ can affect even a magnitude-feature front end and is not guaranteed
to be inert.

Our main diagnostic experiments use FRCRN-SE~\cite{zhao2022frcrn} because it
directly predicts a complex ratio mask, making Eq.~\eqref{eq:projection}
apply without recovery. FRCRN reconstructs $\hat S=MX$ with a complex
CNN--RNN encoder--decoder. To test whether the probe extends beyond this strict
mask output, we also analyze MossFormerGAN-SE, an architecturally distinct
ClearerVoice enhancer built on MossFormer2~\cite{zhao2024mossformer2,zhao2025clearervoice}. It combines a real magnitude-mask path with an
additive complex residual $C$,
\begin{equation}
 \hat S=(M_m|X|)e^{j\angle X}+C.
 \label{eq:moss}
\end{equation}
We recover its effective complex mask $M_{\mathrm{eff}}=\hat S/X$, floor $|X|$ at $10^{-8}$, clip
$|M_{\mathrm{eff}}|$ at 10, and apply Eq.~\eqref{eq:projection}. Because of
this stabilization, $(1,1)$ is the projected full-mask output rather than
necessarily the model's exact native waveform. The comparison tests whether the diagnostic
extends beyond a strict complex-ratio-mask output to other mask-based enhancers.

\section{Experimental setup}
\label{sec:setup}

\textbf{Data.} We use the VoiceBank+DEMAND test set
\cite{botinhao2016investigating,thiemann2013diverse}. Its 824 utterances cover
2.5, 7.5, 12.5, and 17.5~dB; we add one remixed version of every utterance at
0, $-2.5$, $-5$, or $-7.5$~dB using the same recorded DEMAND noises to represent a more adverse acoustic
environment. The merged evaluation contains 1,648 signals and 12,422 reference words. As an
acoustically distinct second set, we use 800 WSJ0 utterances convolved with
simulated room impulse responses (target RT60 0.2--0.6~s) and mixed with
recorded WHAM! ambient noise~\cite{wichern2019wham}. The SNR is sampled
uniformly from the continuous interval $[-5,15]$~dB.

\textbf{Systems and protocol.} The detailed diagnosis uses frozen FRCRN-SE;
the second enhancer is MossFormerGAN-SE (Moss). Both use pretrained
ClearerVoice-Studio checkpoints~\cite{zhao2025clearervoice} and run at 16~kHz. The two primary recognizers deliberately contrast their front ends.
Whisper large-v3~\cite{radford2023whisper} is an encoder--decoder that consumes
log-Mel spectrograms. wav2vec~2.0-large-960h-lv60-self
\cite{baevski2020wav2vec2} combines a convolutional waveform encoder with a
Transformer and CTC fine-tuning. We also include a public robust
wav2vec~2.0 checkpoint~\cite{hsu2021robust}, which keeps the same raw-waveform
CTC while using a robustness-oriented training recipe.

Magnitude sweeps use a 0.05 step for the primary FRCRN VoiceBank experiment
and 0.1 steps for the MossFormerGAN and WSJ sweeps. On VoiceBank, phase sweeps vary
$\gamma$ in 0.1 steps at selected magnitude settings: each recognizer's
measured magnitude minimum and the full-magnitude setting for FRCRN, and the
corresponding measured minima for MossFormerGAN. On WSJ, we test the two
measured minima and the full-magnitude setting with $\gamma\in\{0,.5,1\}$.
WER is micro-averaged over reference words. Confidence intervals use paired
bootstrap resampling ($B=10{,}000$), clustering the two VoiceBank versions of
each source utterance. We use measured grid minima to summarize each sweep;
choosing a deployment operating point would require separate validation. Each
$(0,0)$ value is an identity-mask reconstruction through the corresponding
enhancer pipeline, so small differences from the raw noisy WER reflect the
analysis/synthesis path.

\section{Diagnostic results}
\label{sec:results}

\subsection{Prevalence of the SE--ASR mismatch}
\begin{table}[t]
\caption{The motivating SE--ASR mismatch on VoiceBank+DEMAND, averaged across SNRs. MossFormerGAN-SE and FRCRN-SE use ClearerVoice checkpoints; the remaining enhancement systems are trained on the VoiceBank+DEMAND training set, with Simple mag. mask as a classical baseline. All enhancers improve the three objective enhancement metrics over noisy speech, while only MossFormerGAN-SE improves overall Whisper WER.}
\label{tab:prevalence}
\centering
\setlength{\tabcolsep}{2.4pt}
\begin{tabular}{@{}lrrrr@{}}
\toprule
Input & SI-SDR$\uparrow$ & PESQ$\uparrow$ & STOI$\uparrow$ & WER$\downarrow$\\
\midrule
Noisy speech       &  2.4 & 1.61 & .87 &  6.54\\
MossFormerGAN-SE~\cite{zhao2024mossformer2,zhao2025clearervoice}   & 17.5 & 2.95 & .93 & \cellcolor{green!9} 6.05\\
FRCRN-SE~\cite{zhao2022frcrn}           & 16.6 & 2.73 & .93 & \cellcolor{red!5} 6.78\\
MetricGAN+~\cite{fu2021metricgan+}         &  6.3 & 2.77 & .89 & \cellcolor{red!13}10.59\\
Simple mag. mask   & 14.4 & 2.12 & .89 & \cellcolor{red!17}12.09\\
MeanFlowSE~\cite{li2026meanflowse}         & 17.3 & 2.43 & .92 & \cellcolor{red!9} 8.30\\
FlowSE~\cite{wang2025flowse}             & 16.3 & 2.65 & .92 & \cellcolor{red!10} 8.83\\
\bottomrule
\end{tabular}
\end{table}

We evaluate the systems in Table~\ref{tab:prevalence} on
VoiceBank+DEMAND with Whisper large-v3. The table shows the motivating
mismatch. Every
enhancer improves SI-SDR, PESQ, and STOI over the noisy input, yet WER usually
moves in the opposite direction, with MossFormerGAN-SE the only model improving
overall WER. This mismatch motivates the polar-projection diagnostic.

\subsection{Magnitude strength is recognizer-dependent}

Figure~\ref{fig:alpha} shows the central diagnostic result. For
wav2vec~2.0, increasing the FRCRN magnitude correction produces a large WER
drop and then a broad plateau near full magnitude. A paired bootstrap
comparison confirms that the observed improvement over the identity
reconstruction is not a sampling artifact. Whisper behaves differently: its
lowest measured WER occurs under a much weaker magnitude correction, and the
full correction at $\alpha=1$ returns essentially to its own identity-reconstruction
baseline. Put directly, the same sequence of resynthesized waveforms is fed to
both frozen recognizers, but their WER minima occur in different regions of
the $\alpha$ path. This difference is the observable recognizer dependence
that the probe is designed to expose.

This pattern is the main reason we frame polar projection as a diagnostic. A
single explanation based only on weak or strong enhancement would predict a
similar preferred region for both recognizers. The measured curves instead
show that the same enhanced-signal direction can be useful for one ASR model
and nearly neutral, or even undesirable, for another.

\subsection{Estimated phase gives no consistent gain}

\begin{figure}[t!]
  \centering
  \includegraphics[width=\columnwidth]{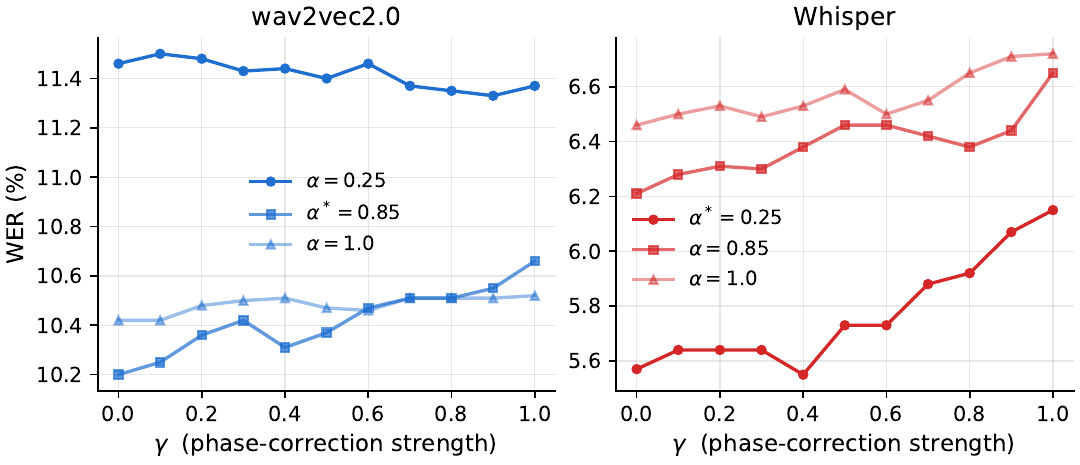}
  \caption{FRCRN phase sweeps at three fixed magnitude settings. The effect of
  $\gamma$ is smaller than the main magnitude response; applying the full
  estimated phase provides no consistent WER gain.}
  \label{fig:gamma}
\end{figure}

Figure~\ref{fig:gamma} shows a different behavior for the FRCRN phase
direction. At full magnitude, increasing $\gamma$ from mixture phase to the full
estimated phase does not produce a reliable improvement for either recognizer. At the
magnitude settings preferred by the two recognizers, using the full estimated
phase ($\gamma=1$) moves WER in the harmful direction. The modified spectrum is resynthesized before recognition: Whisper re-analyzes
that waveform into log-Mel features, while wav2vec~2.0 consumes the waveform
directly. The intervention therefore measures the ASR effect of these
enhancers' \emph{estimated corrections} after magnitude strength is fixed.

MossFormerGAN gives a more nuanced view than the FRCRN slices alone. In point
estimates for wav2vec~2.0, adding the full estimated phase helps at the weaker
$\alpha=0.3$ slice, but not near its $\gamma=0$ magnitude optimum at
$\alpha=0.6$. Whisper does not show a stable benefit from adding the estimated
phase. Across the evaluated slices, estimated phase behaves as a
configuration-dependent effect rather than a stable source of ASR gain.

\subsection{Breadth across systems and acoustic conditions}

\textbf{A second enhancer.}
To test whether the diagnosis depends on FRCRN's complex-mask form, we apply
it to MossFormerGAN through the effective mask in Eq.~\eqref{eq:moss}. The recognizer-dependent magnitude pattern persists:
wav2vec~2.0 favors a strong correction region, while Whisper reaches its best
measured point under weaker modification and then worsens at full magnitude.
The response is therefore observed beyond FRCRN, with enhancer-specific
quantitative differences summarized in Table~\ref{tab:scope}.

\textbf{A third recognizer.}
Table~\ref{tab:scope} also adds a robustness-oriented wav2vec~2.0 checkpoint.
On FRCRN, it prefers a weaker magnitude region than the standard wav2vec~2.0
checkpoint; on MossFormerGAN the two checkpoints are closer. Both recognizers
use raw-waveform CTC models, so input type alone does not determine the
response.

\textbf{A second acoustic test.}
On WSJ0+RIR+WHAM!, wav2vec~2.0 again benefits from a strong FRCRN magnitude
correction. Whisper shows little benefit from the best tested magnitude value,
and the full correction at $\alpha=1$ is worse than its identity
reconstruction. Comparing $\gamma=0$ with $\gamma=1$ at each recognizer's
measured magnitude minimum and at full magnitude follows the same overall
pattern: the estimated phase correction is not a stable source of WER
reduction. The result broadens the evidence to simulated reverberation and
recorded noise.

\section{Discussion and limitations}
\label{sec:discussion}

The main implication of this polar projection diagnosis is that speech enhancement for ASR should be treated as a
recognizer-specific operating-point problem. Enhancement metrics answer whether
the signal is cleaner by conventional criteria; they do not identify how far a
frozen recognizer should move along an enhancer's correction path. Polar
projection makes that operating point explicit by measuring sensitivity along
the predicted magnitude and phase directions, with the enhancer and recognizer
held fixed.

This is directly useful for deployed voice AI and spoken agent pipelines, where
a denoiser, recognizer, and acoustic domain may each be updated independently.
The sweep can serve as a lightweight acceptance test: calibrate $\alpha$ on a
small labeled validation set for the target recognizer, then repeat the same
test when the recognizer, enhancer, or domain changes. A per-utterance oracle
uses reference transcripts in hindsight to choose the best $\alpha$ separately
for each utterance, so it is a lower bound rather than a deployment recipe. Its
gap from the best fixed $\alpha$ shows that different utterances prefer
different enhancement strengths: on the primary FRCRN/VoiceBank setting, WER
falls from 10.20 to 7.84 for wav2vec~2.0 and from 5.57 to 3.74 for Whisper.
This motivates future adaptive $\alpha$ prediction from acoustic or ASR-side
signals~\cite{narayanan2022maskscalar,sato2022learning}.

The phase result should be read in the same pipeline-specific way. Phase
modeling remains valuable for enhancement quality, and prior work shows that
phase can improve perceptual quality and sometimes recognition. Our claim is
more targeted: after resynthesis and recognition, the estimated phase
corrections supplied by the tested enhancers did not provide a stable ASR
advantage once magnitude strength was fixed. The result is also meaningful for log-Mel ASR: the intervention changes the
waveform before Whisper extracts log-Mel features, and the same pattern is
tested with raw-waveform wav2vec~2.0.

The study covers English read speech, two pretrained enhancers, three
recognizer checkpoints, and selected phase slices. Within that evidence base,
the method acts as a compact pipeline audit: it cannot create corrections
absent from the enhancer output, but it can reveal whether the corrections
already present should be used fully, weakened, or separated before recognition.

\begin{table}[t!]
\caption{Cross-system comparison of the preferred point on the magnitude-only
path. Each row reports WER for identity $(0,0)$ $\rightarrow$ best grid point
$\rightarrow$ full magnitude $(1,0)$. VB=VoiceBank+DEMAND;
WSJ=WSJ0+RIR+WHAM!; w2v2=wav2vec~2.0; R=robust checkpoint.}
\label{tab:scope}
\centering
\scriptsize
\renewcommand{\arraystretch}{1.06}
\begin{tabular*}{\columnwidth}{@{\extracolsep{\fill}}lllcr@{\,$\rightarrow$\,}r@{\,$\rightarrow$\,}r@{}}
\toprule
Data & Enhancer & ASR & $\alpha^\star$ & \multicolumn{3}{c@{}}{WER path}\\
\cmidrule(l){5-7}
 & & & & $(0,0)$ & best & $(1,0)$\\
\midrule
VB & FRCRN & w2v2 & .85 & 17.13 & \textbf{10.20} & 10.42\\
VB & FRCRN & w2v2-R & .60 & 17.20 & \textbf{11.39} & 11.60\\
VB & FRCRN & Whisper & .25 & 6.44 & \textbf{5.57} & 6.46\\
VB & MossFormerGAN & w2v2 & .60 & 17.15 & \textbf{9.43} & 9.56\\
VB & MossFormerGAN & w2v2-R & .60 & 17.11 & \textbf{10.55} & 10.84\\
VB & MossFormerGAN & Whisper & .30 & 6.50 & \textbf{5.27} & 5.69\\
WSJ & FRCRN & w2v2 & .70 & 56.52 & \textbf{51.61} & 51.71\\
WSJ & FRCRN & Whisper & .10 & 24.02 & \textbf{23.97} & 25.80\\
\bottomrule
\end{tabular*}
\end{table}

\section{Conclusion}

Inference-time polar projection turns the SE--ASR mismatch into a measurable
pipeline diagnostic. Starting from an enhancer's complex or effective mask, it
controls magnitude strength and estimated phase correction without retraining
either model. This builds on strength-control and observation-adding ideas, but
makes the intervention factorized: instead of only asking whether less
enhancement helps, it identifies whether ASR error follows magnitude
modification, estimated phase correction, or their interaction.

The experiments show that this distinction matters. On VoiceBank+DEMAND,
standard enhancement metrics improve even when Whisper WER often worsens. With
FRCRN, wav2vec~2.0 favors strong magnitude correction, while Whisper reaches
its best WER under a much weaker correction. The same recognizer-dependent
pattern appears with a second enhancer and a second acoustic test, and the
robustness-oriented wav2vec~2.0 checkpoint shows that the preferred operating
point can shift within the same raw-waveform ASR family.

The phase result is meaningful even for log-Mel ASR: projection changes the
waveform before Whisper extracts log-Mel features, and wav2vec~2.0 consumes the
waveform directly. Under this intervention, magnitude strength explains the
larger ASR response, while estimated phase gives no consistent benefit across
the evaluated configurations. SE front ends for voice AI and spoken agents
should therefore be validated at the pipeline level, using measured ASR
behavior to decide whether to use the predicted correction fully, weaken it, or
separate magnitude from estimated phase.

\section{Acknowledgments}

This work used the Delta system at the National Center for Supercomputing
Applications through allocation CIS250338 from the Advanced Cyberinfrastructure
Coordination Ecosystem: Services \& Support (ACCESS) program~\cite{boerner2023access},
which is supported by U.S. National Science Foundation grants \#2138259,
\#2138286, \#2138307, \#2137603, and \#2138296.

\begingroup
\small
\renewcommand{\baselinestretch}{.80}\selectfont
\bibliographystyle{IEEEbib}
\bibliography{refs}
\endgroup

\end{document}